# ACCESSIBILITY ANALYSIS OF SOME INDIAN EDUCATIONAL WEB PORTALS

## Manas Ranjan Patra[1] and Amar Ranjan Dash[2]

[1]Professor, Department of Computer Science, Berhampur University, India
[2]Research Scholar, Department of Computer Science, Berhampur University, India

## ABSTRACT

*Web portals are being considered as excellent means for conducting teaching and learning activities electronically. The number of online services such as course enrollment, tutoring through online course materials, evaluation and even certification through web portals is increasing day by day. However, the effectiveness of an educational web portal depends on its accessibility to a wide range of students irrespective of their age, and physical abilities. Accessibility of web portals largely depends on their user-friendliness in terms of design, contents, assistive features, and online support. In this paper, we have critically analyzed the web portals of thirty Indian Universities of different categories based on the WCAG 2.0 guidelines. The purpose of this study is to point out the deficiencies that are commonly observed in web portals and help web designers to remove such deficiencies from the academic web portals with a view to enhance their accessibility.*

## 1. INTRODUCTION

In recent years, universities across the world have demonstrated an unprecedented level of enthusiasm for their electronic presence by hosting web portals. Web portals are being considered as effective means to reach out to the students by providing user-friendly interfaces. Students also find it convenient to search and access online study materials through web portals. However, the design of web portal is crucial, for making it accessible to different categories of students irrespective of their background and disabilities. The Guidelines proposed by the W3C such as WCAG 2.0, UAAG 2.0, and ATAG 2.0 are being insisted upon to make web portals more accessible.

Since the beginning of web accessibility initiatives in 1997, the researchers have carried out studies to evaluate the accessibility of web portals and user agents. Abdulmohsen Abanumy et al. [1] evaluated the accessibility of government web portals of Saudi Arabia and Oman using WCAG 1.0 guideline. Irina Ceaparu and Ben Shneiderman [2] analyzed the home pages of fifty US State Web sites. J. M. Kuzma et al. [3] examined the accessibility of government web portals of different countries in European Union, Asia, and Africa with respect to the guidelines of WCAG 1.0using TAW as software tool. Similarly, Mustafa Al-Radaideh et al.[4] have evaluated 25 government websites of Jordan using TAW with respect to WCAG 1.0 guideline. Malaysian researchers [5] have used WCAG 1.0 priority 1 and Bobby as an automatic analysis tool to evaluate 9 websites. They have provided only the number of errors of different portals instead of computing the percentage of errors. Thus, it doesn't give an exact picture of the overall error percentage which can be used as a comparative measure among different web portals. Maslina Abdul Aziz et al. [6] have evaluated the usability and accessibility of 120 Malaysian educational websites based on WCAG 1.0. Jeff Carter and Mike Markel [7] have classified disable people into four categories and demonstrate the tools made for them, by different companies. Faouzi Kamoun et al.[8] have made a comparative analysis of web sites of Dubai using both WCAG 1.0 and WCAG 2.0. Manas Ranjan Patra et al. [9] have evaluated the accessibility features of fifteen Indian web portals, both manually and using online tools, with respect to WCAG 2.0 Guideline. They have extended their work by analyzing the accessibility of fifteen government web portals of other Asian countries in [10]. In that work, the accessibility of four embedded media players have also been analyzed using UAAG 2.0. Rita Ismailova and Yavuz Inal [11] have analyzed the accessibility of government websites of four countries viz., Kyrgyzstan, Azerbaijan, Kazakhstan and Turkey. Ibtehal S Baazejem and Hend S. Al-Khalifa[12] have looked into 23 accessibility evaluation studies, and have analyzed the methodologies and tools used during the last five years for analyzing the accessibility of web components.

Abid Ismail and K.S. Kuppusamy [13] have evaluated the accessibility of only the homepages of 302 Indian universities using Achecker on the basis of WCAG 2.0. Tahani Alahmadi and Drew Steve [14] have evaluated the accessibility of websites of top 60 international universities over the period 2005 to 2015. Aidi Ahmi and Rosli Mohamad [15] have evaluated the accessibility of 20 Malaysian university websites based on WCAG 2.0 and Section 508 by using Achecker and WAVE. Hayafa Y. Abuaddous et al. [16] have evaluated the increase in accessibility of 20 public educational universities of Malaysia, from 2012 to 2013. Shaun K. Kane et al.[17] have evaluated the accessibility of homepages of 100 top international universities based on WCAG 1.0.
Till 2004, in most research works the researchers have looked into the accessibility of web components. Then in 2005, Wendy A. Chisholm and Shawn Lawton Henry [18] have classified the web accessibility guideline into three categories WCAG (web component accessibility guideline), ATAG (Authoring Tool Accessibility Guideline), UAAG (User Agent Accessibility Guideline).

They also describe how these three guidelines can control three different aspects of a web portal. Thereafter, researchers have focused on the accessibility features of user agents and authoring tools. Boukari Souley and Amina S. Sambo [19] have analyzed the web browsers namely, Mozilla Firefox, Internet Explorer, Netscape and Opera based on their performance, usability, accessibility and security features. Manas Ranjan Patra and Amar Ranjan Dash [20] have evaluated accessibility of 10 web browsers of Personal Computers based on UAAG 2.0. They have further evaluates the accessibility the accessibility of web browsers of android mobiles in [21]. Jalal Mahmud et al.[22] have designed a context driven non-visual web browser 'CSurf'. Yevgen Borodin et al., [23] have designed 'HearSay', a non-visual web browser with cognitive directed browsing and extensible Voice XML dialog.

The rest of the paper is structured as follows: in section 2, we briefly present the background of web accessibility and the international guidelines. In section 3, we present our methodology for evaluation, and the tools used for accessibility evaluation of web components. In section 4, we evaluate the web portals of some leading academic institutions in India with respect to the International guidelines. In section 5, we analyze the result of our evaluation. Finally, we recommend incorporation of certain features to make the educational web portals more accessible.

## 2. WEB ACCESSIBILITY

Today web portals are major sources of educational information for varieties of students. Thus, it is necessary that, those should be accessible to each and every student, irrespective of age, level of literacy or any physical disabilities. Accessibility of web portals refers to the degree to which it is available to as many students as possible. In order to achieve this, the World Wide Web Consortium (W3C) has launched Web Accessibility Initiative (WAI) in April 1997. The Web Accessibility Initiatives consist of several working groups (Authoring Tool Accessibility Guide-lines Working Group [24], User Agent Accessibility Guidelines Working Group [25], Web

Content Accessibility Guidelines Working Group [26], etc.) each with their specific focus. The working groups have introduced three standardizations to ensure accessibility of web portals at three levels. A schematic representation of the guidelines included in WCAG 2.0, ATAG 2.0, and UAAG 2.0 is presented in Figure 1.

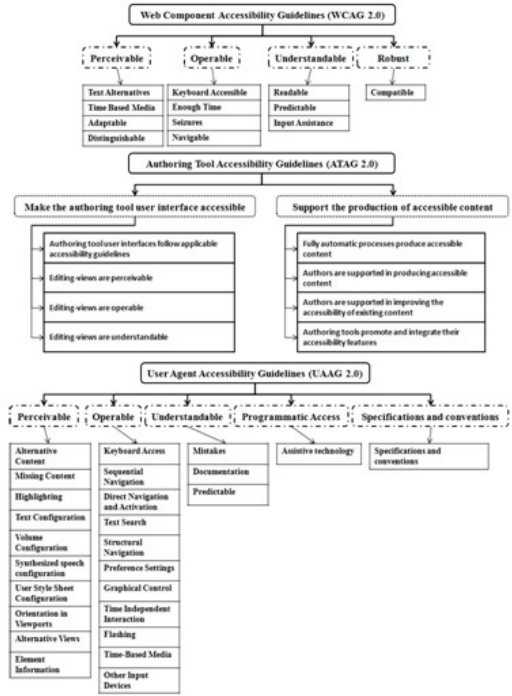

Figure 1. A schematic representation of WCAG 2.0, ATAG 2.0, UAAG 2.0

Web Component Accessibility Guideline [27] has been recommended by the WCAG working group to verify accessibility of web contents. WCAG is an international standard for maintaining and verifying accessibility of the contents of a web portal. The first version of WCAG came out in the form of WCAG 1.0 on 5th May 1999. Subsequently, by adding new features (guidelines needed to be technology-neutral and guidelines needed to be worded as testable) a new version, WCAG 2.0 was recommended by W3C on 11 December 2008. Now, WCAG has been approved by ISO as an international standard (ISO/IEC40500:2012) on October 2012. WCAG 2.0 recommends twelve guidelines based on four basic principles, viz., perceivable, operable, understandable, and robust. A comparative study of WCAG 1.0 and WCAG 2.0 has been done by [28]. Full fledge representation of the WCAG 2.0 guidelines are represented in Figure 2.

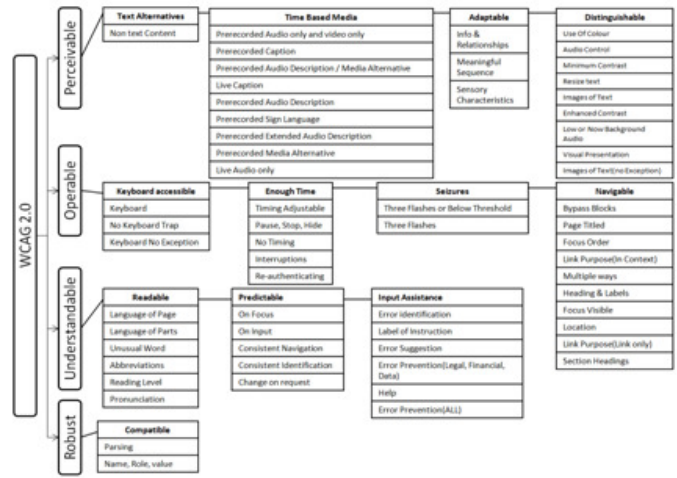

Figure 2.  A schematic representation of WCAG 2.0 Guideline

Authoring Tool Accessibility Guideline [29] has been recommended by the ATAG working group to ensure accessibility of authoring tools. ATAG is an international standard for maintaining and verifying accessibility of authoring tools. The first version of ATAG came out in the form of ATAG 1.0 on 3rd February 2000 and later by adding extra features and removing some of the limitations, a new version of ATAG, ATAG 2.0, was recommended by W3C on 24th September 2015.

User Agent Accessibility Guideline [30] has been recommended by the UAAG working group to verify accessibility of user agents and assistive technologies. UAAG is an international standard for maintaining and verifying accessibility of different user agents (like web browsers, media players) and assistive technologies (like text to speech synthesizer). The first version of UAAG came out in the form of UAAG 1.0 on 17th December 2002 and later by adding extra features and removing some of the limitations a new version is underway. The last working draft of UAAG (UAAG 2.0) was published by W3C on 15TH December 2015 which has 26 guidelines based on five principles, namely, perceivable, operable, understandable, programmatic access, specifications and conventions.

Accessibility of web portals can be controlled and improved in three ways, namely, by following ATAG for the authoring tools during creation of web portals, by controlling the contents of the web using WCAG and by controlling the features of user agent (Web browser and embedded media-player) using UAAG. In the present study we focus on the second aspect, namely, accessibility analysis of components of web portals with respect to WCAG 2.0 recommendations.

## 3. METHODOLOGY

We have considered three different categories of educational websites for our accessibility study by considering a total of thirty educational websites, ten from each category. The features of each of these websites are meticulously studied for compliance where the testing parameters are based on WCAG 2.0 guidelines. The accessibility parameters are checked both manually and by using certain tools as described below. In Figure 3, we represent the overall methodology adopted by us for evaluation of websites.

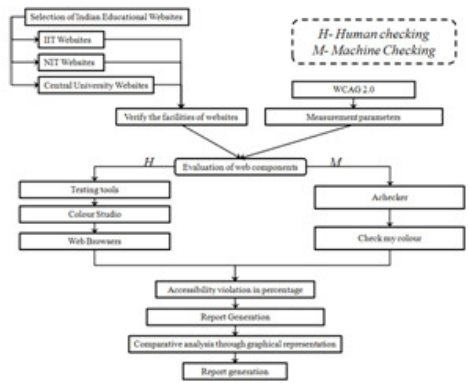

Figure 3. Methodology for accessibility analysis of Educational Web Portals

Table 1 shows some of the tools used in the measurement of different accessibility parameters of the websites. For checking the colour contrast of links we used "Check my colour". For checking the colour contrast of text and images of text we used two tools. In order to obtain the hexadecimal colour code we used "Colour Scheme studio" and then the colour contrast was checked with the help of the online tool "Colour Contrast Check". Different web browsers like Aurora, Comodo Dragon, and Google Chrome were used for accessing the web sites. We also used the "AChecker" tool for checking some of the parameters as per the WCAG 2.0 Guidelines.

Table 1. List of Tools

| Target | Tool used |
|---|---|
| Colour Contrast (link) | Check my colour http://www.checkmycolours.com/ |
| Colour Contrast (image's Text) | Colour Scheme studio http://www.colorschemer.com |
| | Colour contrast check http://webaim.org/resources/contrastchecker/ |
| Web Browser | Mozilla Firefox 52.02 https://www.mozilla.org/en-US/firefox/new |
| | Comodo Dragon 52.15 http://filehippo.com/download_comodo_dragon |
| | Opera 38.0 http://www.opera.com |
| Rule Wise checker | Achecker http://achecker.ca/checker/index.php |

## 4. EVALUATION

This research is intended to provide an accessibility evaluation of 30 Indian educational websites based on WCAG 2.0. For our analysis we have considered 30 educational websites (10 websites each from different IIT, NIT, Central Universities), mentioned in the table. The data taken for our analysis is gathered during the period 10th Sep 2016 to 10th Dec 2016.

Table 2.  Sample Web Portals

| Categories | Universities | Web portals |
|---|---|---|
| IITs | Bombay | www.iitb.ac.in |
| | Delhi | www.iitd.ac.in |
| | Guwahati | www.iitg.ac.in |
| | Indore | www.iiti.ac.in |
| | Jaipur | iitj.ac.in |
| | Kanpur | www.iitk.ac.in |
| | Kharagpur | www.iitkgp.ac.in |
| | Roorkee | www.iitr.ac.in |
| | Varanasi | iitbhu.ac.in |
| | Madras | www.iitm.ac.in |
| NITs | Kurukshetra | www.nitkkr.ac.in |
| | Calicut | www.nitc.ac.in |
| | Bhopal | www.web.manit.ac.in |
| | Rourkela | www.nitrkl.ac.in |
| | Karnataka | www.nitk.ac.in |
| | Tiruchirappalli | www.nitt.edu |
| | Warangal | www.nitw.ac.in/main |
| | Jaipur | www.mnit.ac.in |
| | Allahabad | mnnit.ac.in |
| | Nagpur | www.vnit.ac.in |
| Central Universities | Bihar | www.cusb.ac.in |
| | Gujarat | www.cug.ac.in |
| | Haryana | www.cuh.ac.in |
| | Himachal Pradesh | www.cuhimachal.ac.in |
| | Hyderabad | www.uohyd.ac.in |
| | Jammu | www.cujammu.ac.in |
| | Karnataka | www.cuk.ac.in |
| | Kerala | cukerala.ac.in |
| | Odisha | cuo.ac.in |
| | Tamil nadu | cutn.ac.in |

In order to measure all the dictum of WCAG 2.0 for different components of web portals, first we have checked all the elements and features provided by the web portals and then evaluated the web portals by browsing those using different browsers and platforms. We have documented our findings in a tabular form by indicating the percentage of violations made by web portals against each guideline of WCAG 2.0. "0%" indicates that the web portal does not violate the corresponding guidelines at all whereas "-" indicates that the web components which deal with those guidelines are not present in the web portal. Based on this, the average error percentage is calculated. Figure 4 describes the average percentage violation per rule according to the WCAG 2.0 Guideline.

According to Perceivable principle of WCAG, a web portal should be able to make its component available to human senses. While analyzing the perceivable features of the educational websites, we tested the following component of web portals: "multimedia components", "alternative text for image and videos", "clarity of text based on size and colour", "representation of text through different table and list". During analysis of guidelines under perceivable principle, we found that the guideline which is mostly violated by educational websites is guideline 1.3 (Adaptable) with 45.83% violation. Violations with respect to guideline 1.1 (Text Alternative), guideline 1.2 (Time Based Media), and guideline 1.4 (Distinguishable) were found to be 38.93%, 38.10%, and 20.38% respectively. Although the percentage is less but this rule is violated by approximately all websites, so we provide a detail graphical analysis in Figure 5. It is observed that the top 3 violated sub rules are guideline 1.4.8 (Visual Presentation) with 50.00% violation, 1.4.6 (Enhanced contrast) with 47.45% violation, and guideline 1.4.1 (Use of Colour) with 32.71% violation.

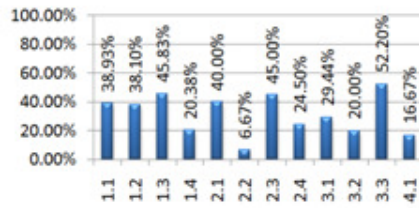

Figure 4.  Violation of accessibility guidelines by the selected educational web Portals

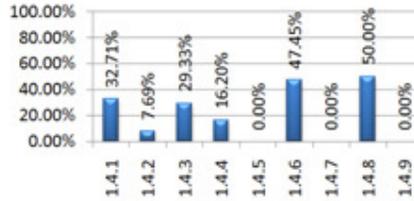

Figure 5.  Violation of accessibility guideline "Distinguishable" by the selected educational web Portals

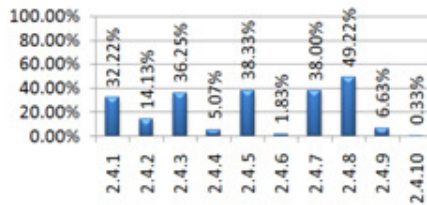

Figure 6.  Violation of accessibility guideline "Navigable" by the selected educational web Portals

According to Operable principle of WCAG, web portals should be able to make their components easily operable by users. While analyzing the operable features of the educational websites, we tested the following components of web portals: "different type of navigation between Components", "Presence of flash, and its features", "keyboard accessibility of components". During analysis of guidelines under operable principle, we found that the guideline which is mostly violated by educational websites is guideline 2.3 (Seizures) with 45.00% violation. But, for guideline 2.1 (Keyboard Accessible), guideline 2.2 (Enough Time), and guideline 2.4 (Navigable) the violation percentages are 40.00%, 6.67%, and 24.50% respectively. Although the percentage is less but this rule is violated by approximately all websites, so we provide a detail graphical analysis in Figure 6. Here we find that the top 3 violated sub rules for navigability are guideline 2.4.8 (Location) with 49.22% violation, guideline 2.4.5 (Multiple ways) with 38.33% violation, and guideline 2.4.7 (Focus Visible) with 38.00% violation.

| | IITs | | | | | | | | | |
|---|---|---|---|---|---|---|---|---|---|---|
| | Bombay | Delhi | Guwahati | Indore | Jodhpur | Kanpur | Kharagpur | Roorkee | Banaras | Madras |
| G.1.1 | 50.00% | 33.33% | 33.33% | 16.67% | 28.57% | 28.57% | 42.86% | 50.00% | 33.33% | 71.43% |
| G.1.2 | - | - | - | - | 50.00% | 50.00% | 16.67% | - | - | 33.33% |
| G.1.3 | 25.00% | 0.00% | 25.00% | 25.00% | 75.00% | 50.00% | 0.00% | 50.00% | 25.00% | 25.00% |
| G.1.4 | 26.15% | 6.20% | 10.75% | 37.63% | 8.82% | 21.24% | 19.52% | 21.71% | 19.34% | 21.05% |
| G.1.4.1 | 17.25% | 9.61% | 27.84% | 63.06% | 4.76% | 38.16% | 5.67% | 55.43% | 5.82% | 29.49% |
| G.1.4.2 | - | - | - | - | 0.00% | 0.00% | 0.00% | - | - | 0.00% |
| G.1.4.3 | 17.95% | 0.00% | 9.96% | 35.00% | 14.81% | 16.00% | 65.00% | 40.00% | 22.22% | 40.00% |
| G.1.4.4 | 60.00% | 0.00% | 0.00% | 0.00% | 5.00% | 5.00% | 0.00% | 0.00% | 0.00% | 0.00% |
| G.1.4.5 | 0.00% | 0.00% | 0.00% | 0.00% | 0.00% | 0.00% | 0.00% | 60.00% | 0.00% | 0.00% |
| G.1.4.6 | 55.83% | 0.00% | 9.96% | 58.00% | 14.81% | 52.00% | 65.00% | 60.00% | 66.67% | 60.00% |
| G.1.4.7 | 0.00% | 0.00% | 0.00% | 0.00% | 0.00% | 0.00% | 0.00% | 0.00% | 0.00% | 0.00% |
| G.1.4.8 | 60.00% | 60.00% | 40.00% | 80.00% | 80.00% | 80.00% | 40.00% | 40.00% | 60.00% | 60.00% |
| G.1.4.9 | 0.00% | 0.00% | 0.00% | 0.00% | 0.00% | 0.00% | 0.00% | 0.00% | 0.00% | 0.00% |
| G.2.1 | 50.00% | 0.00% | 0.00% | 50.00% | 50.00% | 50.00% | 25.00% | 50.00% | 25.00% | 50.00% |
| G.2.2 | 0.00% | 0.00% | 0.00% | 0.00% | 0.00% | 0.00% | 0.00% | 0.00% | 25.00% | 0.00% |
| G.2.3 | 50.00% | 50.00% | 50.00% | 50.00% | 50.00% | 50.00% | 50.00% | 0.00% | 50.00% | 50.00% |
| G.2.4 | 9.72% | 22.03% | 26.60% | 26.57% | 25.26% | 25.18% | 25.30% | 29.40% | 23.33% | 23.51% |
| G.2.4.1 | 0.00% | 33.33% | 0.00% | 33.33% | 33.33% | 33.33% | 33.33% | 33.33% | 33.33% | 33.33% |
| G.2.4.2 | 0.00% | 0.00% | 98.00% | 40.00% | 98.00% | 100.00% | 100.00% | 40.00% | 0.00% | 0.00% |
| G.2.4.3 | 42.18% | 0.00% | 0.00% | 21.95% | 65.46% | 69.97% | 0.00% | 18.70% | 0.00% | 63.48% |
| G.2.4.4 | 0.00% | 0.00% | 3.00% | 15.00% | 0.00% | 15.00% | 18.00% | 4.00% | 0.00% | 5.00% |
| G.2.4.5 | 0.00% | 100.00% | 0.00% | 50.00% | 50.00% | 50.00% | 50.00% | 0.00% | 50.00% | 50.00% |
| G.2.4.6 | 0.00% | 0.00% | 0.00% | 0.00% | 0.00% | 0.00% | 0.00% | 0.00% | 0.00% | 0.00% |
| G.2.4.7 | 40.00% | 0.00% | 14.97% | 9.38% | 5.33% | 23.48% | 46.71% | 85.00% | 45.00% | 16.67% |
| G.2.4.8 | 0.00% | 85.00% | 100.00% | 0.00% | 0.00% | 60.00% | 0.00% | 95.00% | 100.00% | 100.00% |
| G.2.4.9 | 15.00% | 0.00% | 0.00% | 0.00% | 0.00% | 0.00% | 5.00% | 18.00% | 0.00% | 0.00% |
| G.2.4.10 | 0.00% | 0.00% | 0.00% | 0.00% | 0.00% | 0.00% | 0.00% | 0.00% | 0.00% | 0.00% |
| G.3.1 | 0.00% | 0.00% | 50.00% | 50.00% | 50.00% | 16.67% | 16.67% | 16.67% | 16.67% | 33.33% |
| G.3.2 | 0.00% | 0.00% | 40.00% | 40.00% | 0.00% | 40.00% | 40.00% | 40.00% | 40.00% | 40.00% |
| G.3.3 | 57.14% | 14.29% | 57.14% | 57.14% | 57.14% | 71.43% | 71.43% | - | 42.86% | 14.29% |
| G.4.1 | 50.00% | 0.00% | 0.00% | 0.00% | 50.00% | 50.00% | 0.00% | 0.00% | 0.00% | 0.00% |

Figure 7(a). Result of Accessibility Evaluation of IIT web portals

| | NITs | | | | | | | | | |
|---|---|---|---|---|---|---|---|---|---|---|
| | Kurukshetra | Calicut | Bhopal | Rourkela | Surathkal | Tiruchirappalli | Warangal | Malviya | Mesdat Nehru | Visvesvaraya |
| G.1.1 | 66.67% | 66.67% | 50.00% | 33.33% | 16.67% | 50.00% | 50.00% | 14.29% | 28.57% | 42.86% |
| G.1.2 | - | - | - | - | - | - | - | 33.33% | 50.00% | 66.67% |
| G.1.3 | 75.00% | 25.00% | 75.00% | 25.00% | 75.00% | 25.00% | 50.00% | 25.00% | 75.00% | 25.00% |
| G.1.4 | 17.50% | 25.57% | 13.87% | 28.39% | 30.57% | 8.35% | 20.31% | 24.47% | 12.68% | 20.00% |
| G.1.4.1 | 60.00% | 4.58% | 43.21% | 29.78% | 36.63% | 26.78% | 47.02% | 26.88% | 39.14% | 20.68% |
| G.1.4.2 | 0.00% | 48.00% | - | 23.53% | 45.45% | 0.00% | - | 0.00% | 0.00% | 0.00% |
| G.1.4.3 | 0.00% | 12.00% | - | 75.00% | 75.00% | 0.00% | 14.29% | 66.67% | 15.00% | 33.33% |
| G.1.4.4 | 0.00% | 0.00% | 0.00% | 75.00% | 75.00% | 0.00% | 0.00% | 0.00% | 0.00% | 0.00% |
| G.1.4.5 | 0.00% | 0.00% | 15.00% | 0.00% | 0.00% | 0.00% | 0.00% | 0.00% | 0.00% | 0.00% |
| G.1.4.6 | 0.00% | 60.00% | 0.00% | 58.82% | 67.50% | 0.00% | 57.14% | 66.67% | 0.00% | 66.67% |
| G.1.4.7 | 0.00% | 0.00% | 0.00% | 0.00% | 0.00% | 40.00% | 0.00% | 0.00% | 0.00% | 0.00% |
| G.1.4.8 | 80.00% | 80.00% | 40.00% | 40.00% | 20.00% | 0.00% | 40.00% | 60.00% | 60.00% | 60.00% |
| G.1.4.9 | 0.00% | 0.00% | 0.00% | 0.00% | 0.00% | 0.00% | 0.00% | 0.00% | 0.00% | 0.00% |
| G.2.1 | 50.00% | 0.00% | 50.00% | 50.00% | 50.00% | 50.00% | 50.00% | 50.00% | 50.00% | 0.00% |
| G.2.2 | 0.00% | 25.00% | 0.00% | 25.00% | 0.00% | 25.00% | 0.00% | 0.00% | 25.00% | 0.00% |
| G.2.3 | 50.00% | 50.00% | 50.00% | 50.00% | 50.00% | 50.00% | 50.00% | 50.00% | 0.00% | 0.00% |
| G.2.4 | 22.84% | 33.63% | 29.27% | 22.70% | 12.45% | 14.46% | 33.40% | 16.00% | 6.39% | 9.30% |
| G.2.4.1 | 33.33% | 33.33% | 0.00% | 0.00% | 0.00% | 33.33% | 0.00% | 33.33% | 33.33% | 0.00% |
| G.2.4.2 | 0.00% | 100.00% | 100.00% | 100.00% | 0.00% | 0.00% | 100.00% | 100.00% | 100.00% | 0.00% |
| G.2.4.3 | 82.10% | 0.00% | 73.17% | 70.00% | 59.16% | 28.36% | 55.68% | 16.67% | 41.89% | 0.00% |
| G.2.4.4 | 15.00% | 0.00% | 0.00% | 4.00% | 0.00% | 0.00% | 18.00% | 0.00% | 15.00% | 8.00% |
| G.2.4.5 | 0.00% | 100.00% | 100.00% | 0.00% | 50.00% | 0.00% | 100.00% | 0.00% | 0.00% | 0.00% |
| G.2.4.6 | 98.00% | 15.00% | 0.00% | 98.00% | 0.00% | 14.00% | 0.00% | 0.00% | 0.00% | 0.00% |
| G.2.4.7 | 0.00% | 0.00% | 4.55% | 40.00% | 15.38% | 0.00% | 1.28% | 0.00% | 6.98% | 85.00% |
| G.2.4.8 | 0.00% | 80.00% | 100.00% | 15.00% | 0.00% | 45.00% | 55.00% | 0.00% | 0.00% | 0.00% |
| G.2.4.9 | 0.00% | 8.00% | 8.00% | 0.00% | 0.00% | 24.00% | 4.00% | 10.00% | 10.00% | 0.00% |
| G.2.4.10 | 0.00% | 0.00% | 0.00% | 0.00% | 0.00% | 0.00% | 0.00% | 0.00% | 0.00% | 0.00% |
| G.3.1 | 50.00% | 16.67% | 16.67% | 33.33% | 33.33% | 33.33% | 33.33% | 16.67% | 16.67% | 50.00% |
| G.3.2 | 0.00% | 40.00% | 0.00% | 0.00% | 0.00% | 0.00% | 0.00% | 40.00% | 40.00% | 40.00% |
| G.3.3 | 57.14% | 57.14% | 71.43% | 42.86% | 57.14% | 100.00% | 14.29% | 0.00% | 42.86% | 14.29% |
| G.4.1 | 50.00% | 50.00% | 0.00% | 0.00% | 50.00% | 50.00% | 50.00% | 50.00% | 0.00% | 0.00% |

Figure 7(b). Result of Accessibility Evaluation of NIT web portals

According to Understandable principle of WCAG, web browser should be able to make its contents, interface, documentation, and websites component easily understood by users. While analyzing the understandable features of the educational websites, we tested the following components of web portals: "Predictability of components", "readability of text", "features of online forms". During analysis of guidelines under operable principle, we found that the guideline which is mostly violated by educational websites is guideline 3.3 (Input Assistance) with 52.20% violation. Educational websites indicate 29.44% and 20.00% violation in guideline 3.1 (Readable) and Guideline 3.2 (Predictable) respectively. According to our analysis, the Indian Educational Websites have 10.00% violation in accordance to Guideline 4 (Robust). Figure 7(a), 7(b), 7(c) depicts the result of accessibility evaluation of individual web portals.

| | Central University Sites | | | | | | | | | | |
| | Bihar | Gujarat | Haryana | Himachal Pradesh | Hyderabad | Jammu | Karnatk | Kerala | Orissa | Tamilnadu | Avg |
|---|---|---|---|---|---|---|---|---|---|---|---|
| G1.1 | 28.57% | 60.00% | 50.00% | 75.00% | 0.00% | 50.00% | 20.00% | 42.86% | 20.00% | 14.29% | 38.93% |
| G1.2 | 50.00% | 33.33% | | 28.57% | - | - | 50.00% | 16.67% | - | 16.67% | 38.10% |
| G1.3 | 50.00% | 50.00% | 50.00% | 50.00% | 50.00% | 50.00% | 50.00% | 75.00% | 75.00% | 50.00% | 43.83% |
| G1.4 | 50.60% | 26.46% | 17.92% | 25.59% | 10.82% | 20.18% | 19.66% | 30.01% | 29.49% | 22.48% | 20.38% |
| G1.4.1 | 47.64% | 28.14% | 33.33% | 13.48% | 11.56% | 71.43% | 36.00% | 45.07% | 23.77% | 78.14% | 32.71% |
| G1.4.2 | 0.00% | 0.00% | 0.00% | 100.00% | 0.00% | 0.00% | 0.00% | 0.00% | 0.00% | 0.00% | 7.69% |
| G1.4.3 | 88.89% | 30.00% | 0.00% | 25.00% | 5.00% | 0.00% | 0.00% | 75.00% | 69.56% | 25.81% | 29.35% |
| G1.4.4 | 10.00% | 80.00% | 0.00% | 0.00% | 0.00% | 50.00% | 0.00% | 0.00% | 0.00% | 30.00% | 16.20% |
| G1.4.5 | 0.00% | 0.00% | 0.00% | 0.00% | 0.00% | 0.00% | 0.00% | 0.00% | 0.00% | 0.00% | 0.00% |
| G1.4.6 | 88.89% | 80.00% | 50.00% | 50.00% | 10.00% | 60.00% | 60.00% | 90.00% | 82.60% | 48.39% | 47.45% |
| G1.4.7 | 0.00% | 0.00% | 0.00% | 0.00% | 0.00% | 0.00% | 0.00% | 0.00% | 0.00% | 0.00% | 0.00% |
| G1.4.8 | 40.00% | 20.00% | 0.00% | 40.00% | 60.00% | 40.00% | 40.00% | 60.00% | 60.00% | 20.00% | 50.00% |
| G1.4.9 | 0.00% | 0.00% | 0.00% | 0.00% | 0.00% | 0.00% | 0.00% | 0.00% | 0.00% | 0.00% | 0.00% |
| G2.1 | 50.00% | 25.00% | 50.00% | 25.00% | 50.00% | 75.00% | 50.00% | 50.00% | 50.00% | 25.00% | 40.00% |
| G2.2 | 25.00% | 0.00% | 25.00% | 0.00% | 0.00% | 0.00% | 0.00% | 25.00% | 0.00% | 0.00% | 6.67% |
| G2.3 | 50.00% | 50.00% | 33.47% | 50.00% | 50.00% | 50.00% | 50.00% | 50.00% | 50.00% | 50.00% | 45.00% |
| G2.4 | 18.78% | 29.35% | | 26.85% | 17.74% | 47.60% | 25.55% | 22.48% | 36.98% | 31.80% | 24.50% |
| G2.4.1 | 33.33% | 100.00% | 66.67% | 66.67% | 0.00% | 100.00% | 66.67% | 33.33% | 66.67% | 33.33% | 32.22% |
| G2.4.2 | 0.00% | 0.00% | 100.00% | 100.00% | 100.00% | 100.00% | 0.00% | 0.00% | 100.00% | 100.00% | 33.57% |
| G2.4.3 | 52.97% | 0.00% | 26.00% | 0.00% | 79.37% | 0.00% | 48.41% | 34.04% | 43.00% | 35.03% | 36.25% |
| G2.4.4 | 0.00% | 5.00% | 0.00% | 10.00% | 0.00% | 0.00% | 0.00% | 10.00% | 0.00% | 5.00% | 5.07% |
| G2.4.5 | 0.00% | 0.00% | 50.00% | 0.00% | 0.00% | 0.00% | 0.00% | 0.00% | 100.00% | 100.00% | 38.33% |
| G2.4.6 | 0.00% | 23.88% | 0.00% | 0.00% | 0.00% | 10.00% | 5.00% | 0.00% | 10.00% | 0.00% | 3.10% |
| G2.4.7 | 27.59% | 19.67% | 2.00% | 91.87% | 98.00% | 98.00% | 7.69% | 97.42% | 3.51% | 98.00% | 38.00% |
| G2.4.8 | 58.93% | 35.00% | 80.00% | 60.00% | 40.00% | 98.00% | 78.00% | 0.00% | 46.67% | 95.00% | 49.22% |
| G2.4.9 | 15.00% | 0.00% | 0.00% | 0.00% | 10.00% | 10.00% | 0.00% | 0.00% | 0.00% | 35.00% | 6.63% |
| G2.4.10 | 0.00% | 0.00% | 0.00% | 0.00% | 0.00% | 0.00% | 0.00% | 0.00% | 0.00% | 0.00% | 0.33% |
| G3.1 | 16.67% | 50.00% | 50.00% | 50.00% | 16.67% | 16.67% | 50.00% | 33.33% | 16.67% | 33.33% | 29.44% |
| G3.2 | 40.00% | 0.00% | 40.00% | 0.00% | 40.00% | 0.00% | 0.00% | 0.00% | 0.00% | 40.00% | 20.00% |
| G3.3 | 42.86% | 71.43% | 71.43% | | 57.14% | 57.14% | | 85.71% | | 71.43% | 52.20% |
| G4.1 | 0.00% | 0.00% | 0.00% | 0.00% | 0.00% | 0.00% | 0.00% | 50.00% | 0.00% | 0.00% | 16.67% |

Figure 7(c).  Result of Accessibility evaluation of Central University web portals

# 5. RESULT ANALYSIS

Figure 8 provides analysis of the accessibility violation done by websites of each category with respect to the four principle of WCAG 2.0. Based on this data, we have calculated the percentage of accessibility violation done by web portals of selected categories with respect to different principles. Figure 9 provides a comparative view of guideline violations by the three selected categories of websites with respect to the criteria - perceivable, operable, understandable, and robustness. With respect to the Perceivable principle most violation is found in NIT websites (37.02%) and least violation is found in IIT Websites (28.63%). According to the rule of "text alternative" NIT, IIT, and Central university websites show 41.90%, 38.81% & 36.07% respectively. In case of "adaptable" criterion Central University websites (57.50%) indicate more violation than NIT websites (50.00%) and IIT websites (30.00%).

| | IIT Violation | | NIT Violation | | CU Violation | | Total |
|---|---|---|---|---|---|---|---|
| | Percentage | Avg | Percentage | Avg | Percentage | Avg | |
| Perceivable | 31.11% | | 40.23% | | 37.32% | | 108.66% |
| | | 28.63% | | 37.02% | | 34.35% | |
| Operable | 26.64% | | 27.51% | | 32.97% | | 87.13% |
| | | 30.58% | | 31.58% | | 37.85% | |
| Understandable | 32.43% | | 30.57% | | 38.21% | | 101.21% |
| | | 32.04% | | 30.20% | | 37.76% | |
| Robust | 15.00% | | 30.00% | | 5.00% | | 50.00% |
| | | 30.0% | | 60.0% | | 10.0% | |

Figure 8.  Accessibility violation of principles by each category

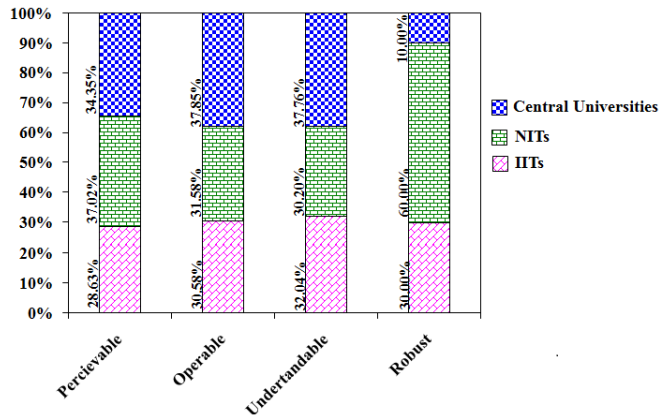

Figure 9.  Percentage wise Accessibility violation of principles by each category

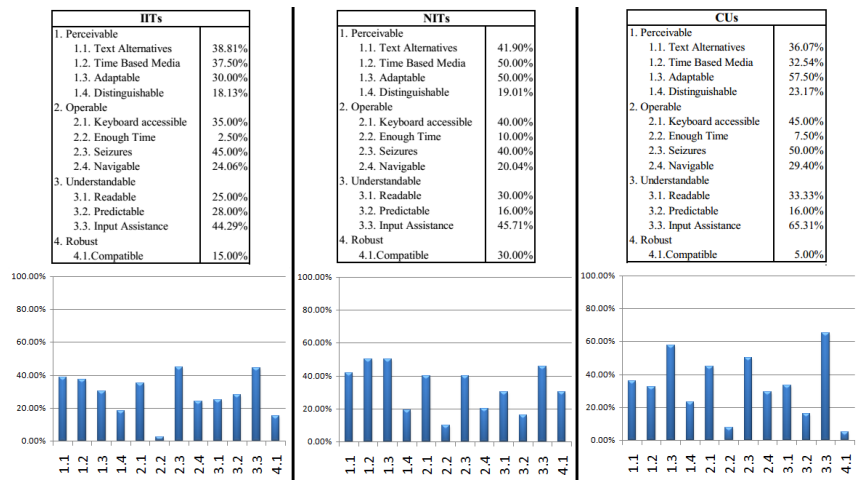

Figure 10.  Accessibility Analysis of individual Categories

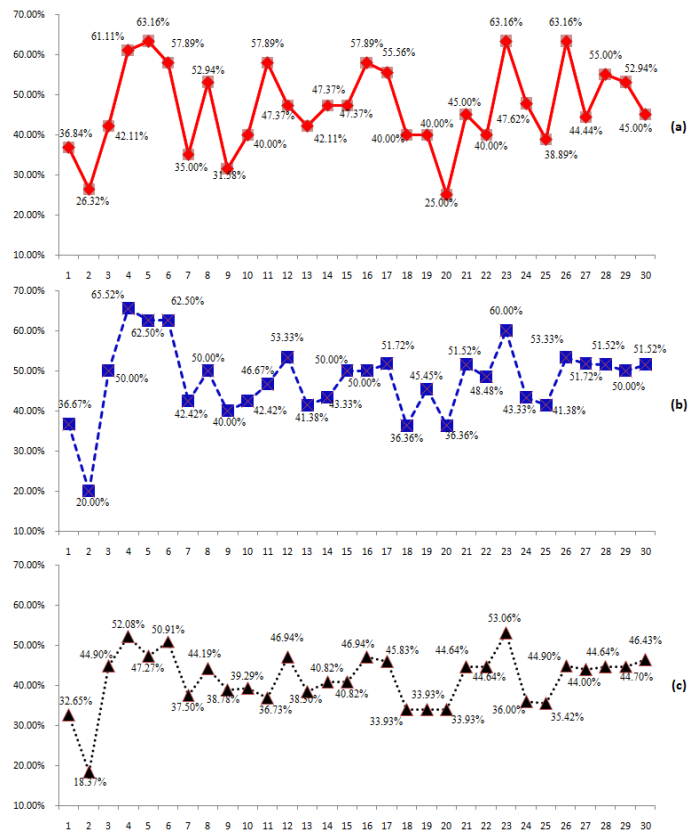

Figure 11. Accessibility violation of different web portals with respect to (a) conformance level A (b) conformance level AA (c) conformance level AAA

In case of the Operable Principle most violation is found in Central University websites (37.85%) and least violation is found in IIT Websites (30.58%). According to the keyboard accessibility criteria, Central University, NIT, and IIT websites show 45.00%, 40.00%, and 35.00% violations respectively. According to the "location" criteria, IIT websites (64.00%) and Central University websites (51.66%) show more violation than NIT websites (32.00%). In case of "multiple ways" guidelines IIT websites show 45% violation, whereas both NIT and Central University show 35%. According to this criterion a website should contain any two of "A list of related pages", "Table of contents", "Site map", "Site search", and "List of all available Web Pages".

The "Understandability" criterion is mostly violated by Central University websites with 37.76% violation and least violated by NIT Websites with 30.20%. According to the "Input Assistance" criterion, Central University, NIT, and IIT websites indicate 65.31%, 45.71% and 44.29% violations respectively. From our analysis we found that the online forms available in the IIT Delhi website is much more accessible compared to other educational websites. As per the Robustness guidelines most violation is found in NIT websites (60%) and least violation is found in Central University Websites (10%). A detail analysis under each category is provided in Figure 10.

WCAG 2.0 consists of four principles, twelve guidelines, and three conformance levels. According to Level-A Conformance, a web portal must satisfy a minimum set of requirements to make it accessible to normal people. According to Level-AA Conformance, a webpage should satisfy some additional features including those in Level-A to make it accessible to disable persons. Similarly, according to Level-AAA, a webpage should satisfy each guideline of WCAG (including all three priority levels) to make it harmlessly accessible to all. Figure 11 depicts the Accessibility violation of different web portals with respect to different conformance levels.

## 6. RECOMMENDATIONS

Based on our study the following recommendations are made with respect to the design of web portals:

- Text alternatives should be provided for all non-text web contents,
- Headers to be included for each page, section, and table,
- There should be provision for controlling colour contrast and support for all keyboard functionalities, and
- All the forms must be well structured and interactive in nature.

## 7. CONCLUSION

With the growing number of web portals worldwide, accessibility issues have emerged as a serious concern for web designers. In this paper, we have evaluated the accessibility of three major categories of Indian educational web portals and their compliance with respect to the WCAG 2.0 guidelines. It is observed that most of the websites do not conform to all the criteria as laid down by the guidelines in order that they are accessible to all. In particular, the average accessibility violations by the selected web browsers range from 11.44% (IIT Delhi) to 41.17% (Central University, Haryana). We believe that the quantitative results of our evaluation can help web designers of educational institutes to incorporate the required features according to the WCAG 2.0 guidelines in order to make web portals more pragmatic and accessible to various user categories.

## AUTHORS

Dr. Manas Ranjan Patra holds a Ph.D. Degree in Computer Science from the Central University of Hyderabad. He is a Professor in the Department of Computer Science, Berhampur University, India and has been teaching Computer Science for the last 29 years. He was a United Nations Fellow at IIST/UNU, Macao. He has more than 150 research publications to his credit. His research interests include Service Oriented Computing, Applications of Data mining and e-Governance. He has extensively travelled to many countries for presenting research papers, chairing technical sessions and delivering invited talks. He has been a member of Editorial Boards and Programme Committees of many International journals and conferences. He is a member of ACM, CSI and ISTE. 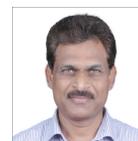

Mr. Amar Ranjan Dash has obtained his B. Tech. degree from Biju Patnaik University of Technology, India and M. Tech. degree from Berhampur University, India. Currently, he is pursuing his doctoral research. He has 5 research publications to his credit. His research interests include Web Accessibility, Web Service Choreography, and Cloud Computing. He is a member of ACM. 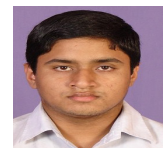